\documentclass[journal,twocolumn]{IEEEtran}
\usepackage{epsfig,makeidx,color}
\usepackage{amsmath,amssymb,bbm}
\usepackage{cite,graphicx,lipsum}
\usepackage{enumerate}
\usepackage[switch,pagewise]{lineno}
\usepackage{hyperref}
\hypersetup{
        colorlinks = true,
        citecolor=blue,
}
\def\rT{{\rm T}}

\def\uE{{\mathbb E}}

\DeclareMathOperator*{\argmax}{\arg\!\max}

\def\be{ \begin{equation} }
\def\ee{ \end{equation} }
\def\bea{ \begin{eqnarray} }
\def\eea{ \end{eqnarray} }

\def\bx{{\bf x}}

\def\ba{{\bf a}}

\def\bu{{\bf u}}

\def\bw{{\bf w}}

\def\bI{{\bf I}}

\def\b0{{\bf 0}}

\def\cD{{\cal D}}

\def\cN{{\cal N}}

\def\cW{{\cal W}}

\ifCLASSOPTIONonecolumn
  \newcommand{\figwidth}{0.38\columnwidth}
  
\else
  \newcommand{\figwidth}{0.87\columnwidth}
  
\fi

\begin{document}

\title{Federated Learning with Multichannel ALOHA}

\author{Jinho Choi and Shiva Raj Pokhrel\\
\thanks{The authors are with
the School of Information Technology,
Deakin University, Geelong, VIC 3220, Australia
(e-mail: \{jinho.choi, shiva.pokhrel\}@deakin.edu.au).}}


\maketitle
\begin{abstract}
In this paper, we study federated learning in a cellular system 
with a base station (BS) and a large number of users with local 
data sets.
We show that multichannel random access can provide
a better performance than sequential polling
when some users are unable to compute local updates 
(due to other tasks) or in dormant state.
In addition, for better aggregation in federated learning,
the access probabilities of users can be optimized for given
local updates. To this end, we formulate an optimization problem
and show that a distributed approach
can be used within federated learning to adaptively
decide the access probabilities.
\end{abstract}


\ifCLASSOPTIONonecolumn
\baselineskip 22.5pt
\fi

\section{Introduction}

Federated learning 
\cite{FO16, Yang19, Park19} has been extensively
studied as a distributed machine learning
approach with data privacy.
In federated learning, 
mobile phones or devices
keep their data sets and exchange
a parameter vector
to be optimized in a certain learning problem
(with data sets that are kept at devices
or users). Throughout the paper, we
interchangeably use devices and users.

Since each user uploads its local update
to a server 
and the server sends the aggregated
update back to users, in cellular setting,
we can assume that all the exchanges are carried out
through base stations (BSs).
In \cite{Zhu18, Amiri19, KYang19},
federated learning in a cell is considered,
where all the users are located in a cell and
communicate with a BS.
Under this setting,
in \cite{Zhu18, KYang19}, the notion of over-the-air
computation \cite{Nazer07, Goldenbaum13}
is adopted for aggregation when analog versions
of local updates are transmitted, where
the BS receives a noisy version of the aggregated
update. In \cite{Amiri19},
fading channels are taken into account.

In this paper, we consider the setting that
all the users are located in a cell and a BS
is to communicate with them
as in \cite{Zhu18, Amiri19, KYang19}.
However, we do not consider the notion of over-the-air
computation.
For efficient uploading with a limited system bandwidth,
we consider multichannel random access (e.g.,
multichannel ALOHA) \cite{Chang15, Choi16CL}.
In most cases for federated learning,
it is assumed that all the users
are able to upload their local updates at each iteration,
and sequential polling with multiple 
access channels can allow to upload more local updates
at the cost of wider bandwidth.
However, in practice,
some users (e.g., mobile phones) might be busy for other tasks
or are in dormant state.
As a result, although they are asked to upload,
no local updates from them are available,
which results in waste of channels.
This motivates us to use multichannel random 
access, i.e., multichannel ALOHA, for
more efficient uploading than polling in federated learning.

Furthermore, for effective aggregation 
with a small number of local updates,
the access probability of each user in multichannel ALOHA
can be optimized.
To this end, we formulate an optimization problem and
find the solution with a distributed implementation method
in conjunction with federated learning.

\section{System Model}

Suppose that a federated learning
system (FLS) consists of $K$ mobile devices and one BS
in a cell
as in \cite{Zhu18, Amiri19, KYang19}.
As mentioned earlier, we use mobile devices and 
user interchangeably.
In FLS, each user has its data set $\cD_k = \{\bx_k, y_k\}$,
where $\bx_k$ and $y_k$ 
represent the input and output of user $k$,
respectively \cite{FO16},
and there is a parameter or weight vector $\bw$ associated with 
the following optimization problem:
$\min_\bw \frac{1}{K} \sum_{k=1}^K f_k (\bw)$,
where
$f_k (\bw)$ denotes the loss function at user $k$, which has in general
the following form:
$f_k (\bw) = \ell(\bw, \bx_k, y_k)$.
For example, the loss function for linear regression is given by
\be
\ell(\bw, \bx_k, y_k) = \frac{1}{2} | \bx_k^\rT \bw - y_k|^2.
	\label{EQ:ell}
\ee

In FLS, the users do not upload their data set to the BS, but
send their local updates for given weight vector through iterations.
Let $\bw (t)$ denote the weight vector at iteration $t$,
where $t$ is the index for the iteration.
Then, user $k$ can find its local update 
with $\cD_k$ as follows:
\be
\bw_k (t+1) \leftarrow {\rm LocalUpdate}(\bx_k, y_k, \bw (t)),
	\label{EQ:wt0}
\ee
where the local update depends on the loss function. For 
the loss function at user $k$ in \eqref{EQ:ell},
the local update
becomes $\bw(t) - h_k (\bx_k^\rT \bw (t) - y_k) \bx_k$ with step size 
$h_k > 0$ for the gradient descent (GD) algorithm. 
Here, $\bw_k (t)$ and $\bw(t)$
represent the weight vector $\bw$ at user $k$ and BS at iteration $t$,
respectively (the subscript $t$ is used for the user index).
Once all the users send their local updates to the BS,
the BS is able to update the weight vector
as follows:
\be
\bw (t+1) = \frac{1}{K} \sum_{k=1}^K \bw_k (t+1),
	\label{EQ:wt1}
\ee
which is referred to as aggregation.

\section{Uploading via Multiple Channels}

In FLS, local updating and aggregation at the BS
in \eqref{EQ:wt0} and \eqref{EQ:wt1}, respectively,
are to be carried out iteratively.
This iteration requires uploading the 
local weight vectors from $K$ users.
If $K$ is large, the required time for uploading per iteration
might be long. To shorten the uploading time,
multiple channels can be used with a wider system bandwidth.
In this section, we consider a random sampling approach
to approximate for \eqref{EQ:wt1} with multiple channels and
combine it with ALOHA.

\subsection{Multichannel Random Access}

The averaging in \eqref{EQ:wt1}
requires all the $K$ local updates. 
Provided that the system bandwidth is 
limited, all the local updates may not be available
at each iteration.
Thus, suppose that there are $M$ parallel channels,
where $M \ll K$ so that $M$ users can upload
their local updates simultaneously at each iteration.

It is possible that the BS chooses
a set of $M$ users to upload their local updates
at each iteration.
Alternatively, a pre-determined user sequence
can be used. In this case, it is not necessary for the BS
to choose $M$ users at each iteration.
However, there are drawbacks. 
First, there can be users that cannot upload their local updates
due to various reasons. For example, 
a user (or sensor) may not be available 
as it is in dormant state,
or its local computation
to find its update cannot be carried out as it is busy with
some other tasks.
Secondly, a user with negligible local update
can be asked to upload its local update, which leads to
a negligible impact on the aggregation in
\eqref{EQ:wt1}.
To avoid the above drawbacks, we can consider multichannel ALOHA
with the access
probability that depends on the local update,
which will be considered in Subsection~\ref{SS:AAP}.

To address the first drawback,
let $p_{\rm comp}$ denote the probability that
a user is able to compute its local update.
Thus, if $M$ users are asked to upload 
their local updates by the BS, only $p_{\rm comp} M$ users on average
are able to send their updates.
As a result, polling with $M$ channels may not be
efficient if $p_{\rm comp}$ is not high 
(as $(1-p_{\rm comp})M$ channels would be 
idle on average).
To overcome this problem, we can consider
multichannel ALOHA, where each user with local update
can randomly choose one of $M$ channels 
with a certain access probability
and sends
its local update. 

Suppose that a user that can compute its local update
can randomly choose one of $M$ channels. 
We assume that the BS cannot receive any local
updates if multiple users choose the same channel
(due to packet collision).
Let $p$ denote the access probability,
i.e., the probability that a user sends its local update.
Clearly, $p \le p_{\rm comp}$.
Then,
the average number of local updates
at the BS
is given by
\begin{align}
\eta = K p \left( 1 - \frac{p}{M} \right)^{K-1} 
\approx Kp e^{- \frac{p K}{M}}  \le M e^{-1}.
\end{align}
The maximum average number of local updates 
can be achieved if $p = \frac{M}{K}$.
As a result, if $p_{\rm comp} \le e^{-1}$ and $M \le e^{-1} K$,
we can see that multichannel ALOHA
(with $p = \frac{M}{K}$)
can upload more local updates than polling on average
for the aggregation in \eqref{EQ:wt1}
as
$\eta \approx M e^{-1} \ge M p_{\rm comp}$.
Note that when multichannel ALOHA is used
with $M$ channels, the access probability becomes
\be
p = \min \left\{\frac{M}{K}, p_{\rm comp}\right\}.
	\label{EQ:p_ma}
\ee

\subsection{Adaptive Access Probability based on Local Update}
\label{SS:AAP}

In this subsection, we address the second drawback.
To this end, we need to allow that the access probability
of each user, which was assumed to be the same for all users
in the previous subsection,
is now different and depends on its local update.

In this subsection, we first formulate an optimization
problem to approximate the aggregation
in terms of the access probabilities of users (from the BS's perspective).
Then, we show that each user can decide its access probability
with its local update and (simple) feedback information from the BS.
Let
$\ba = \sum_{k=1}^K \bw_k$,
which is the unnormalized aggregation.
In addition, define
$\bu = \sum_{k=1}^K \bw_k \delta_k$,
where $\delta_k \in \{0,1\}$ becomes 1 if the BS receives
the local update from user $k$ and 0 otherwise.
We assume that $\delta_k$ is dependent on $\bw_k$.
It can be seen that $\bu$ is 
an approximation of $\ba$ 
for the aggregation
in \eqref{EQ:wt1}.
To see the approximation error, we can 
consider the following conditional error norm:
\begin{align}
\uE[||\ba - \bu|| \,\bigl|\, \cW] & = 
\uE[ ||\sum_{k=1}^K \bw_{k} (1 - \delta_k) || \,\bigl|\, \cW] \cr
& \le \sum_{k=1}^K a_k \uE[ 1- \delta_k \, |\, \bw_k] 
\le \sum_{k=1}^K a_k e^{- q_k},
	\label{EQ:bounds}
\end{align}
where the first inequality is due to 
the triangle inequality
and the second inequality is due to $1 - x \le e^{-x}$ for $x \in (0,1)$,
and $a_k = ||\bw_k||$.
Here, $q_k = \uE[ \delta_k\,|\, \bw_k]$,
which is the probability that the BS
receives the local update from user $k$.

Let $p_k$ denote the probability
that user $k$ transmits its local update
(or the access probability of user $k$).
Then, 
the probability that the BS
successfully
receives the local update from user $k$,
$q_k$, is given by
\begin{align}
q_k = p_k \prod_{n \ne k} \left(1  - \frac{p_n}{M} \right) 
\le p_k e^{ - \sum_{n \ne k} \frac{p_n}{M}}
\le p_k e^{- \frac{P}{M}},
	\label{EQ:qp}
\end{align}
where $P = \sum_{k=1}^K p_k$.
Then, it can be shown that
\begin{align}
Q = \sum_k q_k \le
\sum_k p_k e^{ - \frac{P}{M}}
= P e^{- \frac{P}{M}} \le M e^{-1}.
	\label{EQ:const}
\end{align}
In \eqref{EQ:const}, the second inequality 
becomes the equality
(i.e.,
the sum of the probabilities of successful 
uploadings, $Q$, can be maximized) when $P = M$.
Thus, with $P = M$, from \eqref{EQ:qp},
we can have
$q_k \le p_k e^{-1}$.
If $P = M$, the last inequality becomes
equality in \eqref{EQ:const}. 
Furthermore, since the bounds in \eqref{EQ:qp}
are tight when $K$ is sufficiently large  and $p_n/M$ is sufficiently
low, we will assume 
that
\be
q_k = p_k e^{-1}.
	\label{EQ:qpe}
\ee

From \eqref{EQ:bounds}, \eqref{EQ:const}, and \eqref{EQ:qpe},
$q_k$ can be decided to minimize the error bound
as follows
\begin{eqnarray}
& \min_{q_k} \sum_k a_k e^{- q_k} & \cr
& \mbox{subject to} \ \sum_k q_k = M e^{-1} \ 
\mbox{and} \ q_k \in (0,e^{-1}), \ \forall k,&
	\label{EQ:OP}
\end{eqnarray}
which is a convex optimization problem.
Note that the second constraint, 
$q_k \in (0,e^{-1})$ is due to $p_k \le 1$ and \eqref{EQ:qpe}.
Then, the solution is given by
\be
q_k^* = [\ln a_k - \ln \lambda]_0^{e^{-1}},
	\label{EQ:oq}
\ee
where $\lambda$ is a Lagrange multiplier
and $[x]_a^b$, where $a < b$, is
defined as
$$
[x]_a^b = 
\left\{
\begin{array}{ll}
x, & \mbox{if $a < x < b$} \cr
a, & \mbox{if $x \le a$} \cr
b, & \mbox{if $x \ge b$.} \cr
\end{array}
\right.
$$
In \eqref{EQ:oq}, we can find $\lambda$ to satisfy
$\sum_k q_k^* = M e^{-1}$.

Alternatively, $p_k$ can be obtained from \eqref{EQ:oq}
using \eqref{EQ:qpe}, i.e., $p_k = q_k e$, as follows:
\be
p_k^* = [e \ln a_k - \psi]_0^{1},
	\label{EQ:op}
\ee
where $\psi = e \ln \lambda$.
In addition, in \eqref{EQ:OP}, if a user cannot compute its local update,
it needs to set $a_k = 0$ so that $p_k = 0$.

\subsection{Feedback Signal from the BS}

Finding the solution
of \eqref{EQ:OP} requires all the $a_k$'s. Therefore, 
\eqref{EQ:OP} cannot be carried out at each user.
However, as in \eqref{EQ:op}, if $\psi$
is available at each user, $p_k$ can be found.

Denote by $\hat P_t$ an estimate of $P$ at iteration $t$, which is 
$\hat P_t = \sum_{k=1}^K s_{k ,t}$,
where $s_{k,t} \in \{0,1\}$ represents 
the activity variable of user $k$ at iteration $t$
(i.e., $s_{k,t} = 1$ if user $k$ transmits its local update
at iteration $t$, otherwise
$s_{k,t} = 0$). 
Clearly, $\hat P_k$ is seen as 
the total number of active
users that send their local updates through
$M$ channels (regardless of collisions).
Note that as shown in \cite{Choi16CL}, it is possible to find 
the total number of active users in an existing 
machine-type communication (MTC) standard.

We consider a feedback signal, $\psi$, which is to be sent
from the BS to users at the end of each iteration. 
From \eqref{EQ:const},  using the dual ascent method \cite{Boyd11},
$\psi$ can be adaptively decided to keep $P = M$ close as follows:
\be
\psi_{t+1} = \psi_t + \mu (\hat P_t - M),
	\label{EQ:pmu}
\ee
where 
$\psi_t$ represents the updated $\psi$ at 
iteration $t$ and $\mu$ denotes the step size.
At the end of iteration $t$, the BS can 
send $\bw(t+1)$  as well as  $\psi_{t+1}$
to all the users so that each one
can not only find local updating, but also 
decide whether or not to transmit its local update
according to \eqref{EQ:op}.

Note that the access probability, $p_k$,
is adaptively decided in \eqref{EQ:op}
without knowing $K$ and $p_{\rm comp}$,
which might be another advantage over polling
when $K$ and  $p_{\rm comp}$ are not known to the BS.

\section{Simulation Results}

In this section, the stochastic 
GD (SGD) algorithm
is used in federated learning
with the squared error loss function in \eqref{EQ:ell}.
For simulations,
at each user, we assume that
$\bx_k$ of length $L \times 1$ is an independent Gaussian random vector,
i.e., $\bx_k \sim \cN(\b0, \bI)$.
In addition, in each simulation run, $\bw$ is also generated
as an  independent Gaussian random vector,
i.e., $\bw  \sim \cN(\b0, \bI)$,
while $y_k = \bx_k^\rT \bw$ at user $k$, $k = 1,\ldots, K$.

Prior to presenting the main simulation results,
we demonstrate the importance 
of choosing the users with significant
local updates for uploading when the system bandwidth is limited.
Suppose that only one user can upload at a time (i.e., $M = 1$).
In this case, 
cyclic coordinate descent (CCD) 
as an example of SGD
can be used  as follows \cite{FO16}:
\be
\bw (t+1) = \bw(t) - \mu_1 
(\bx_{k(t)}^\rT \bw(t) - y_{k(t)}) \bx_{k(t)},
	\label{EQ:CCD}
\ee
where $k(t)$ is the user index at iteration $t$
and $\mu_1$ is the step size.
Here, $k(t) = k \ \mbox{mod}\ t$ for CCD.
For simplicity, we assume that $p_{\rm comp} = 1$.
For comparison, another approach
where the user of the largest
local update in terms of its norm
is chosen at each iteration is considered,
where the index of the user 
that is to upload its local update
at iteration $t$ is chosen as
\be
k(t) = \argmax_k
||(\bx_{k(t)}^\rT \bw(t) - y_{k(t)}) \bx_{k(t)}||.
	\label{EQ:kt}
\ee

\begin{figure}[thb]
\begin{center}
\includegraphics[width=\figwidth]{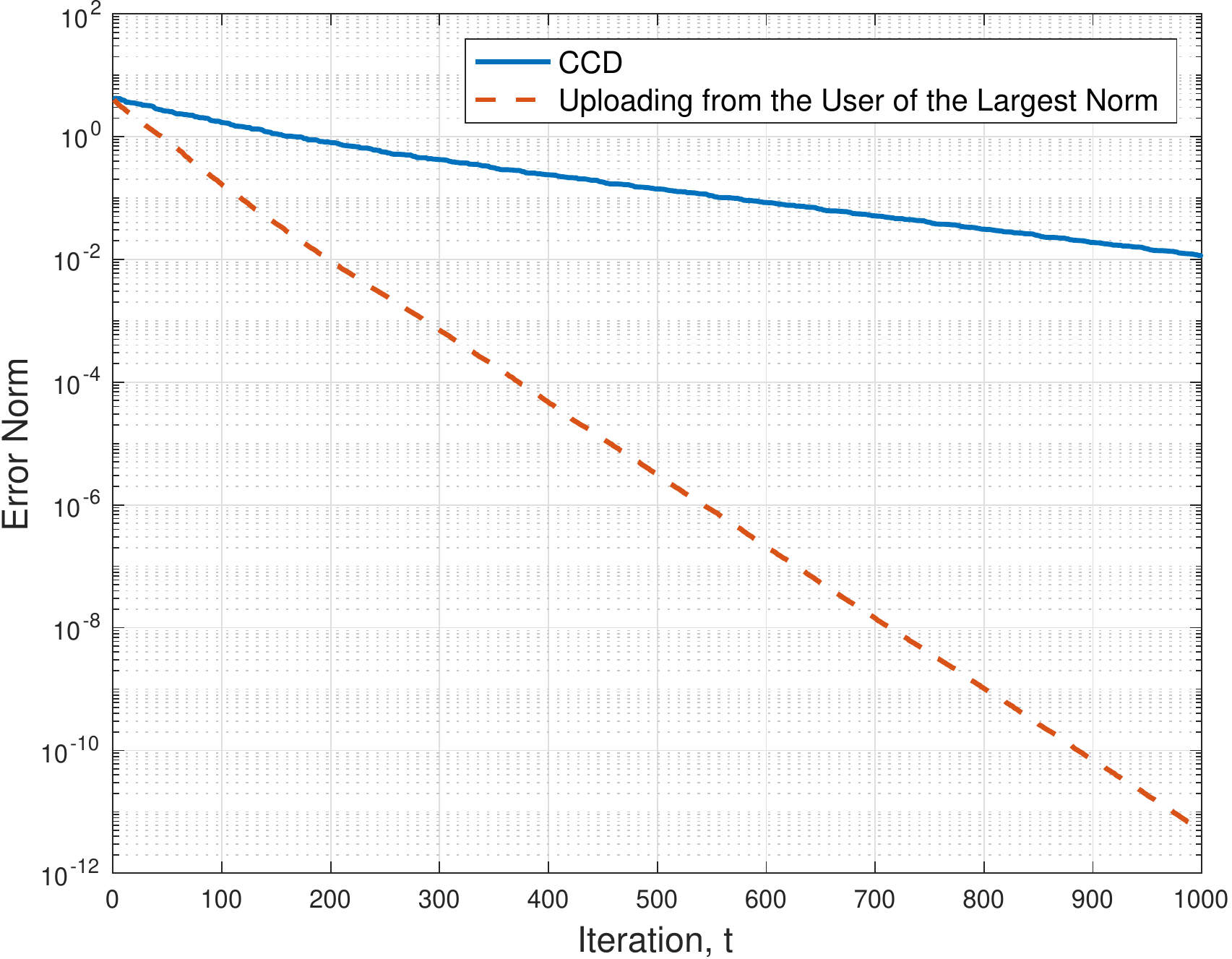} \\
\end{center}
\caption{Error norm, $||\bw (t) - \bw||$, with
CCD and the uploading from the user corresponding to \eqref{EQ:kt}
(at each iteration) is shown as functions of iterations
when $K = 100$, $L = 10$, and $\mu_1 = 0.01$.}
        \label{Fig:plt_rr_ss}
\end{figure}

In Fig.~\ref{Fig:plt_rr_ss},
the error norm,
$||\bw (t) - \bw||$, with
CCD and the uploading from the user corresponding to \eqref{EQ:kt}
(at each iteration) is shown as functions of 
the number of iterations
when $K = 100$, $L = 10$, and $\mu_1 = 0.01$.
Clearly, it is shown that
if the user of the largest
local update (in terms of its norm)
is chosen at each iteration as in \eqref{EQ:kt},
it can significantly improve the performance.
Unfortunately, 
since the user corresponding to
\eqref{EQ:kt} is not known by the BS,
the BS is not able to ask the user to upload
its local update at each iteration.
However, as discussed earlier,
it is possible to take into account the norm
of the local update
when the access probability is decided 
with multichannel ALOHA, which might lead to performance
improvement.

\begin{figure}[thb]
\begin{center}
\includegraphics[width=\figwidth]{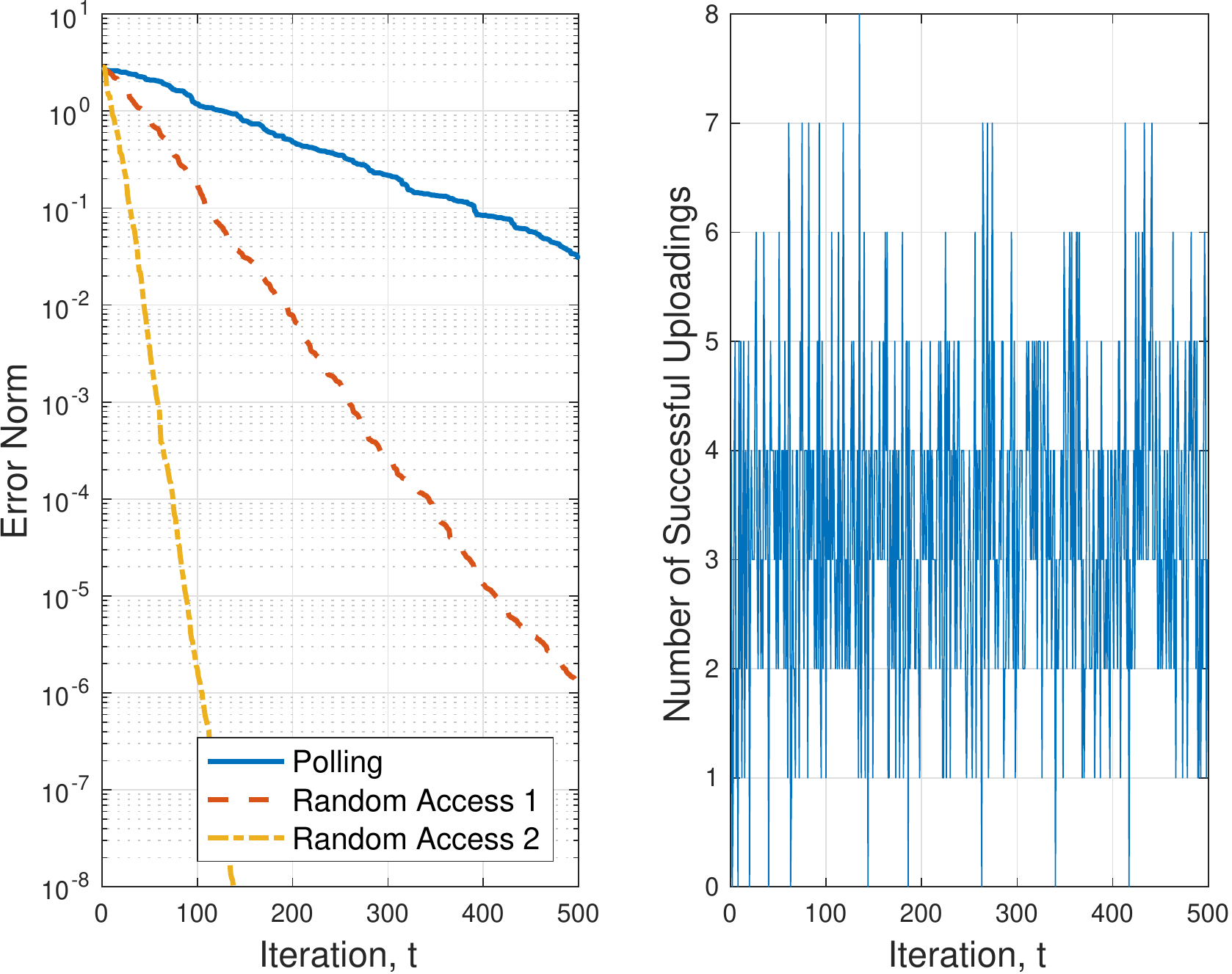} \\
\hskip 0.5cm (a) \hskip 3.5cm (b)
\end{center}
\caption{Performance of three different systems for federated learning
when $K = 1000$, $M = 10$, $L = 10$, $(\mu_1, \mu) = (0.01, 0.1)$,
and $p_{\rm comp} = 0.1$:
(a) Error norm, $||\bw (t) - \bw||$;
(b) Number of successfully uploadings (of Random Access 2).}
        \label{Fig:plt_aa}
\end{figure}

We now consider three different systems. One is based on polling
with $M$ channels (which can be seen as an SGD algorithm
that does not take into account
the significant of local updates (in terms of their norms)
in choosing coordinates)
and the other two systems are based
on multichannel ALOHA.
For convenience, Random Access 1 
denotes the multichannel ALOHA  
system with an equal access probability
of $p$ in \eqref{EQ:p_ma}, while Random Access 2 represents 
the multichannel ALOHA 
system
with the access probability in \eqref{EQ:op}
and \eqref{EQ:pmu}.
In Fig.~\ref{Fig:plt_aa},
we show the 
performance of three different systems for federated learning
when $K = 1000$, $M = 10$, $L = 10$, $(\mu_1, \mu) = (0.01, 0.1)$,
and $p_{\rm comp} = 0.1$.
We can see that Random Access 2 outperforms the others.

\begin{figure}[thb]
\begin{center}
\includegraphics[width=\figwidth]{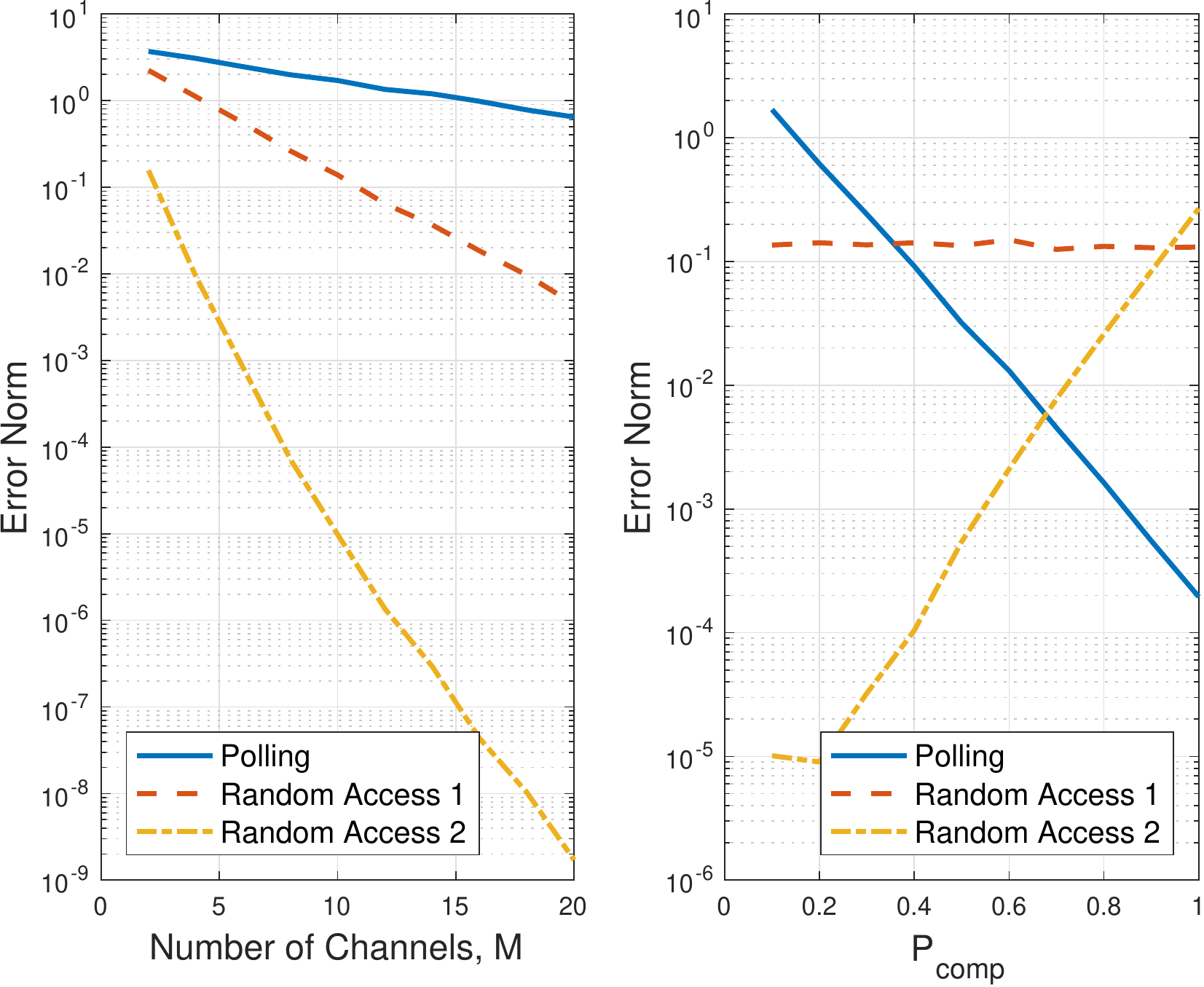} \\
\hskip 0.5cm (a) \hskip 3.5cm (b)
\end{center}
\caption{Error norms of three different systems for federated learning
when $K = 1000$, $L = 10$, $(\mu_1,\mu) = (0.01,0.1)$, and
the number of iterations is set to 100:
(a) Error norm versus $M$;
(b) Error norm versus $p_{\rm comp}$.}
        \label{Fig:plt12}
\end{figure}

In 
Fig.~\ref{Fig:plt12}, the performance
of  three different systems for federated learning
is shown in terms of $M$ and $p_{\rm comp}$
when $K = 1000$, $L = 10$, $(\mu_1,\mu) = (0.01,0.1)$, and
the number of iterations is set to 100.
It is shown in
Fig.~\ref{Fig:plt12} (a), all the systems have improved performance
as $M$ increases.
In Fig.~\ref{Fig:plt12} (b), it is shown that the performance
of Random Access 1 is almost independent of $p_{\rm comp}$,
while its performance is worse than polling when
$p_{\rm comp} > e^{-1}$ as expected.
It is noteworthy that the performance of Random Access 2 
is degraded as $p_{\rm comp}$ increases,
which is due to a convergence time to find $\psi$ in \eqref{EQ:pmu}.

\begin{figure}[thb]
\begin{center}
\includegraphics[width=\figwidth]{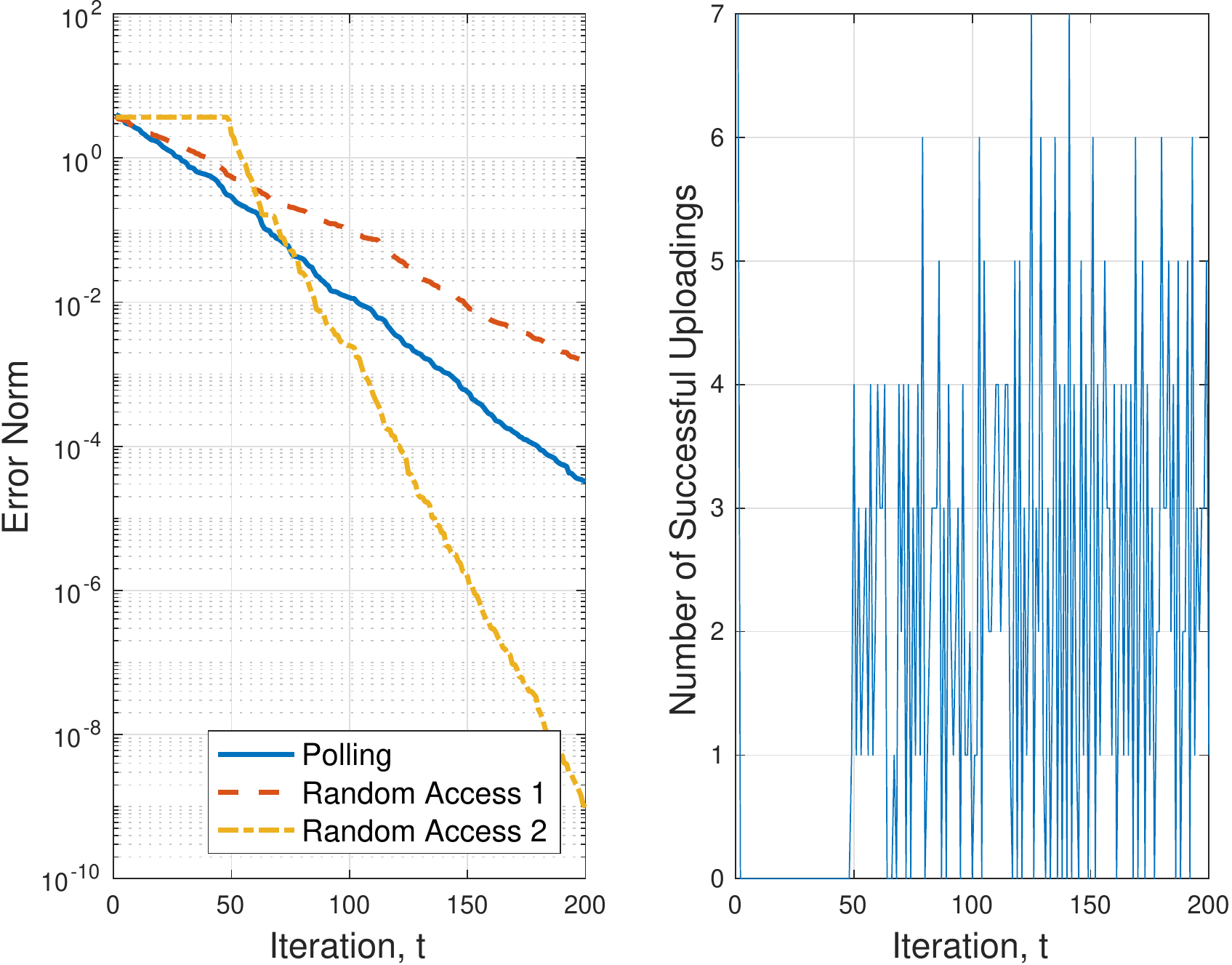} \\
\hskip 0.5cm (a) \hskip 3.5cm (b)
\end{center}
\caption{Performance of three different systems for federated learning
when $K = 1000$, $M = 10$, $L = 10$, $(\mu_1, \mu) = (0.01,0.1)$,
and $p_{\rm comp} = 0.6$:
(a) Error norm, $||\bw (t) - \bw||$;
(b) Number of successfully uploadings (of Random Access 2).}
        \label{Fig:plt_bb}
\end{figure}

In Fig.~\ref{Fig:plt_bb} (a),
we show
the trajectory of error norms
when $K = 1000$, $M = 10$, $L = 10$, $(\mu_1, \mu) = (0.01,0.1)$,
and $p_{\rm comp} = 0.6$.
It is shown that Random Access 2 cannot upload
local updates as $p_k$ is too low for the first 50 iterations,
as shown in Fig.~\ref{Fig:plt_bb} (b).
Once $\psi_t$ becomes low enough through the iteration
in \eqref{EQ:pmu}, $p_k$ becomes sufficiently high
to upload local updates and a better performance can be achieved
with a sufficient number of iterations (say, more than 100 iterations).

\section{Conclusions}

We studied federated learning within a cellular system
and adopted multichannel ALOHA
to upload local updates from a large number of users.
It was shown 
that multichannel ALOHA can perform better than
sequential polling when the probability that a user
is able to upload its local update is less than $e^{-1}
\approx 0.3679$. It was also demonstrated
that the access probability can be optimized with
the significant of local update at each user
(which is measured by the norm of the local update)
for 
better performance in terms of aggregation in federated learning.
A distributed approach for optimizing access probability
was also presented.

\bibliographystyle{ieeetr}
\bibliography{mtc}

\end{document}